\documentclass[prl,aps,nofootinbib,showkeys,showpacs,twocolumn]{revtex4}   
\usepackage{epsfig}  
\usepackage{graphicx}  
\usepackage{amssymb}  
\usepackage{amsmath}
\usepackage{color}  
\begin{document}  
  
\title{Superfluid dynamics of $^{258}$Fm fission}  
  
\author{Guillaume Scamps}  
 \email{scamps@nucl.phys.tohoku.ac.jp}  
\affiliation{GANIL, CEA/DSM and CNRS/IN2P3, Bo\^ite Postale 55027, 14076 Caen Cedex, France} 
\affiliation{Department of Physics, Tohoku University, Sendai 980-8578, Japan}
 \author{C\'edric Simenel}  
 \email{cedric.simenel@anu.edu.au}  
\affiliation{Department of Nuclear Physics, Research School of Physics and Engineering \\ Australian National University, Canberra, Australian Capital Territory 2601, Australia}

\author{Denis Lacroix} \email{lacroix@ipno.in2p3.fr}  
\affiliation{Institut de Physique Nucl\'eaire, IN2P3-CNRS, Universit\'e Paris-Sud, F-91406 Orsay Cedex, France}    
    
\begin{abstract}  
Theoretical description of nuclear fission remains one of the major challenges of  quantum many-body dynamics.
The slow, mostly adiabatic motion through the fission barrier is followed by a fast, non-adiabatic descent of the potential between the fragments. 
The latter stage is essentially unexplored. 
However, it is crucial as it generates most of the excitation energy in the fragments.
The superfluid dynamics in the latter stage of fission is obtained with the time-dependent Hartree-Fock theory including  BCS dynamical pairing correlations. 
The fission modes of the $^{258}$Fm nucleus are studied. 
The resulting fission fragment characteristics  show a good agreement with experimental data. 
Quantum shell effects are shown to play a crucial role in the dynamics and formation of the fragments. 
The importance of quantum fluctuations beyond the independent particle/quasi-particle picture is underlined and qualitatively studied.
\end{abstract}

 
\maketitle

Since its discovery in 1939 \cite{hah39,mei39}, nuclear fission has been a pillar of nuclear physics. 
It is indeed the process revealing most clearly the complexity of low-energy nuclear dynamics. 
Thus, it provides an ideal test for the modelling of nuclear systems with quantum many-body theories. 
It has also found important applications in fundamental science. 
For instance, it is the best way to produce beams of exotic rare isotopes in present and future accelerators.
In addition, understanding fission dynamics is crucial for the production of superheavy elements, an important motivation for the construction of exotic beam accelerators. 
Fission is also present in some astrophysical processes. 
In fact, the natural abundance of elements heavier than iron is believed to be largely influenced by the fission of very neutron-rich heavy nuclei formed in supernovae and neutron star mergers \cite{gor13}. 
Moreover, nuclear fission is one of the greatest sources of energy available on Earth. 
Safely extracting this energy to produce electricity while preserving our environment has been, and still is, one of the greatest technological challenges of humanity.

The theoretical description of the nuclear fission phenomenon remains a profound problem in fundamental science. 
In order for fission to occur, nuclei have to overcome the fission barrier, which involve slow, dissipative motion. 
This process has been widely modelled using the adiabatic approximation \cite{mor91}. 
This approximation is justified near the fission barrier by the fact that the evolution of the collective coordinate is slow enough to allow the internal degrees of freedom to be equilibrated.
However, after the barrier is passed, the fission fragments undergo an accelerating descent to scission, which involves non-adiabatic effects, and so cannot be described with standard adiabatic models. 
Thus, the adiabatic approximation breaks down in the latter stage of fission, where the evolution is faster. 
In particular, the dynamics near scission, where the fragments separate, is clearly non-adiabatic  \cite{riz13,dub08}. 
These effects are crucial to properly describe properties of the fragments such as their mass, charge, and their excitation energy. 
In particular, the latter determines the number of emitted neutrons and is thus one of the most important properties for the simulation and safety of future nuclear reactors. 

The complexity of fission dynamics and the high number of degrees of freedom to be included motivate the use of microscopic approaches, where the quantum behaviour of each and every nucleon in the whole system is followed in time. 
Microscopic approaches have recently considerably improved our understanding of the fission process \cite{gou05,bon06,dub08,sta09,pei09,you11,war12,abu12,mir12,sta13,mcd13,sad13,sim14,sch14a}.
Time dependence is also a key to addressing the latter stage of the fission process \cite{neg78}.
Non-adiabatic effects in the latter stage of fission have then recently been investigated in fission of $^{264}$Fm using a time-dependent mean-field approach \cite{sim14}.
As an example of new outcomes, it was shown that more than half of the final excitation energy of the fragments is acquired during the last zeptosecond before scission and that it is at least partly stored in low-energy collective vibrational states of the fragments.
However, the calculations in Ref.~\cite{sim14} are based on an independent particle approximation and pairing correlations responsible for a superfluid phase in nuclear systems were neglected, thus limiting the range of possible applications to very few non-superfluid systems. 
It is then crucial to incorporate time-dependent pairing correlations in order to investigate fission dynamics across the nuclear chart. 
These correlations have been recently included in realistic time-dependent mean-field calculations \cite{ave08,eba10,ste11,has12,sca13} which we extend to the study of fission.

The purpose of this letter is to present a microscopic method which incorporates both superfluid dynamics and non-adiabatic effects in the latter stage of the fission process.
Fission in the $^{258}$Fm nucleus is considered as an example of application.
This nucleus is known experimentally to exhibit a bimodal fission \cite{hul89} and constitute an ideal benchmark for theoretical studies of fission \cite{mol00,bon06,dub08,ich09,sta09}.

In the present approach, it is assumed that the fission process is divided in two steps. 
In a first step, the slow evolution near the fission barrier is treated in a standard way using the adiabatic approximation. The constrained Hartree-Fock (HF) equations with pairing correlations are solved at the BCS level (CHF+BCS). 
Several constraints are considered in order to find different valleys in the potential energy surface. 
In a second step, the non-adiabatic descent of the potential towards scission is determined using the time-dependent HF equations with dynamical pairing correlations (TDHF+BCS). 
The properties of the fragments, in particular their mass, charge, and kinetic energy, are then computed after scission and compared with experimental data from \cite{hul89}.

 \begin{figure}[!ht]   
	\centering\includegraphics[width=\linewidth]{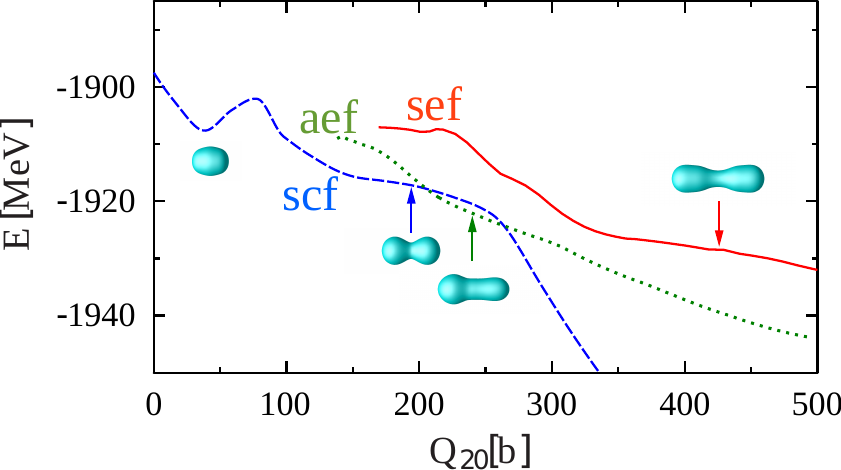}  
	\caption{(Color online) Potential energy in the three valleys: symmetric compact fragment (scf) (blue dashed line), symmetric elongated fragment (sef) (solid red line) and asymmetric elongated fragment (aef) (green doted line). The arrows correspond to the starting configuration of the dynamical calculations for each mode. Isodensities at half the saturation density $\rho_0/2=0.08$~fm$^{-3}$ are also plotted at these initial configurations and for the $^{258}$Fm ground-state (left).} 
	\label{fig:pot} 
\end{figure}

The mean-field is obtained with the Sly4$d$ \cite{kim97} Skyrme energy density functional and a constant-$G$ interaction in the pairing channel. The pairing strength is chosen to be $G_n=24/N$~MeV and $G_p=19/Z$~MeV.
The CHF+BCS and TDHF+BCS calculations are obtained with modified versions of the \textsc{ev8} \cite{bon05} and \textsc{tdhf3d} \cite{kim97} codes, respectively, assuming only one plane of symmetry. 
All calculations are performed on a Cartesian grid of 88$\times$19.2$\times$19.2 fm$^{3}$ with a mesh size 0.8~fm.
The time evolution is obtained with a time step $1.5\times10^{-24}$~s.  

 The CHF+BCS solutions with constraints  on quadrupole $Q_{20}$, octupole $Q_{30}$, and hexadecapole  $Q_{40}$ moments along the fission axis let appear three valleys. 
Two symmetric valleys (i.e., with a total $Q_{30}=0$) lead to symmetric compact fragments (scf) and to symmetric elongated fragments (sef), respectively. 
 In the scf valley, the final fragments are almost spherical while they exhibit a strong prolate shape in the sef valley. 
 A third valley with $Q_{30}\ne0$ leads to asymmetric elongated fragments (aef) with different masses and charges.
Similar adiabatic valleys were obtained by other groups for the same nucleus \cite{bon06,sta09,ich09}. 
The potential energy along these three valleys is shown in Fig. \ref{fig:pot}.
This first, adiabatic, stage of fission is crucial in determining the outcome of the reaction. 
We see that the scf and aef valleys have similar energies up to relatively large deformations ($Q_{20}\sim270$~fm$^2$) in the descent to scission. 
The sef valley, however, is found at higher energy.

\begin{figure}[!ht]   
	\centering\includegraphics[width=\linewidth]{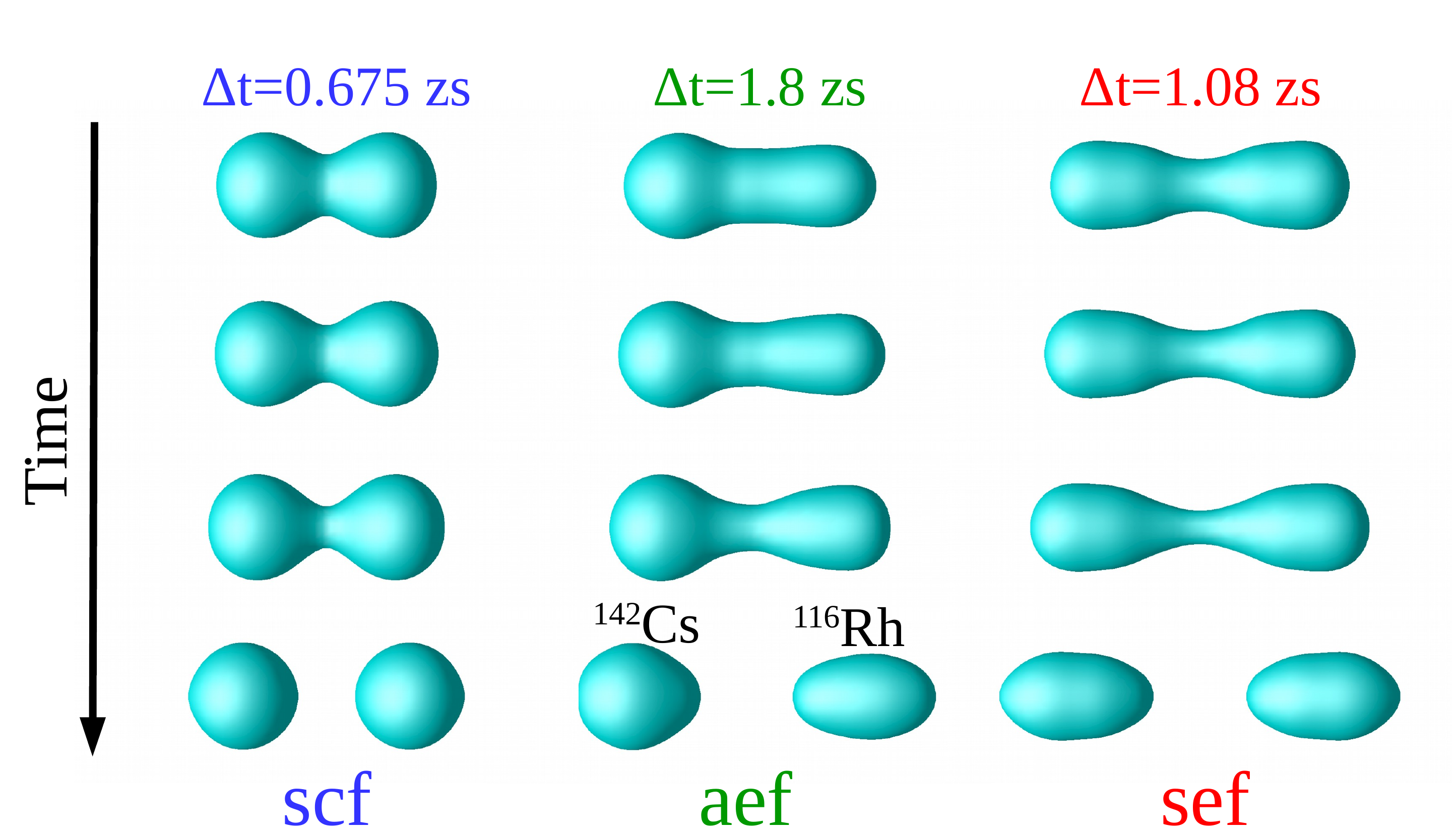}  
	\caption{(Color online) Isodensity surfaces at half the saturation denity $\rho_0/2=0.08$ fm$^{-3}$ as a function of time for the three modes : scf (left), aef (middle) and sef (right). The time step between two images are $\Delta t=0.675$~zs, 1.8~zs and 1.08~zs  for the scf, aef and sef modes, respectively.} 
	\label{fig:film} 
\end{figure}

\begin{figure}[!ht]   
	\centering\includegraphics[width=\linewidth]{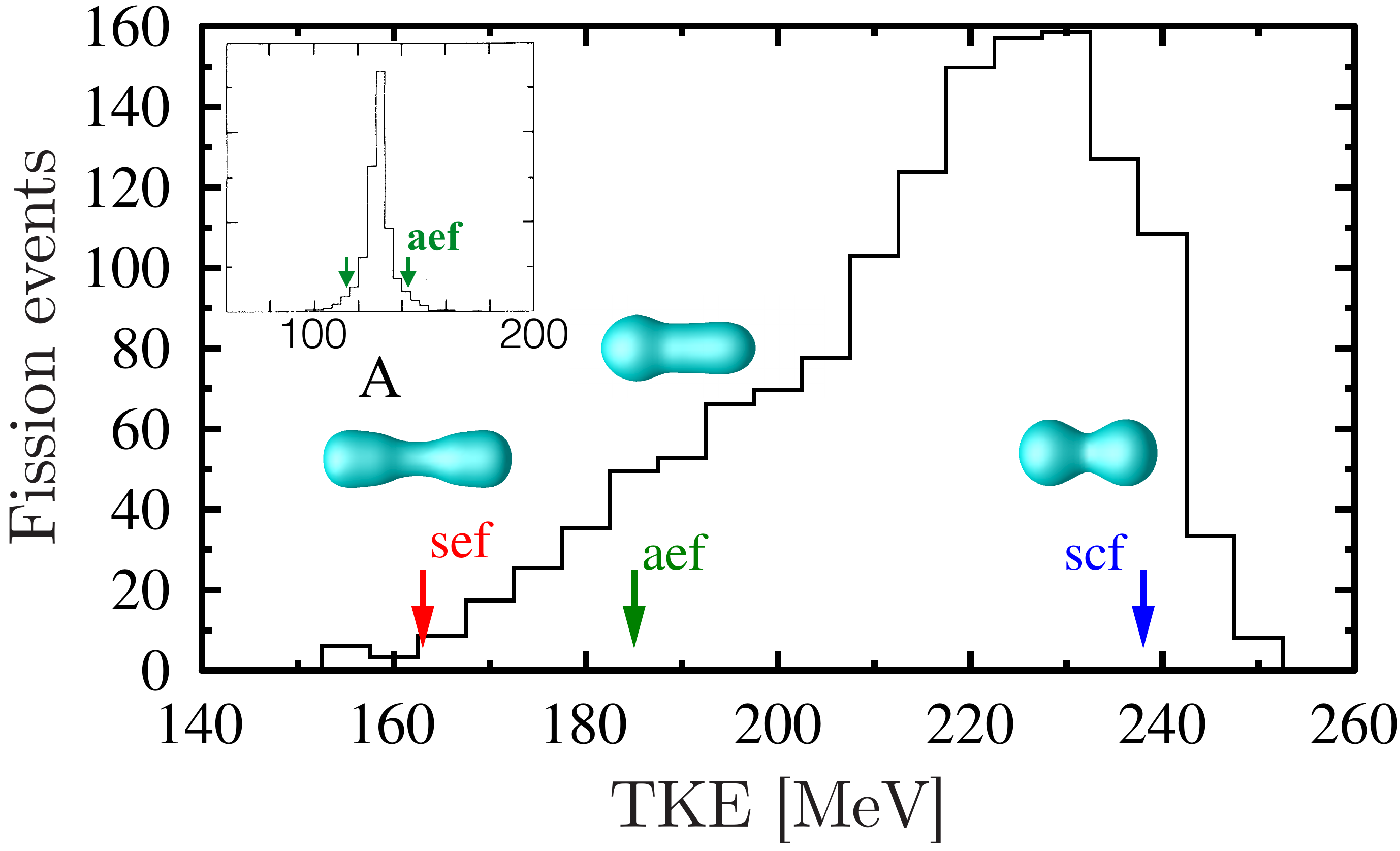}  
	\caption{Experimental distribution of TKE from Ref. \cite{hul89}. The arrows correspond to the mean value of the TKE from the three TDHF+BCS calculations. The fragment mass distribution is shown in the inset. } 
	\label{fig:exp_TKE} 
\end{figure}

Let us now investigate the second stage of the fission process, associated with the non-adiabatic descent of the potential towards scission. 
TDHF+BCS calculations have been performed with initial configurations along theses valleys indicated by arrows in Fig. \ref{fig:pot}. 
More compact configurations belong to the adiabatic phase as single particle states cross the Fermi level and induce a two-body dissipation due to the Landau-Zener effect \cite{blo76,sim14}.
The density evolutions in the non-adiabatic phase are represented at various times for each mode in Fig.~\ref{fig:film}.  
The asymmetric mode is likely to be responsible for the tail of the experimental fragment mass distribution shown in the inset of Fig.~\ref{fig:exp_TKE}.
It is also interesting to note that the three evolutions require different times to reach scission.
These times are $\sim2$~zs, $\sim5.4$~zs, and $\sim3.2$~zs for the scf, aef, and sef modes, respectively. 
These variations are likely to be due to a combination of two factors: different potential slopes and different one-body viscosity which is expected to depend on shell effects.
It is worth mentioning that the dynamical pairing effects significantly affect the fission dynamics. Indeed, freezing pairing correlations by keeping occupation numbers constant in time, the so-called Frozen Occupation Approximation (FOA) \cite{Mar05,sca13}, inhibits strongly the fission process. 

The final total kinetic energy (TKE) of the fragments is another important observable which can be used to distinguish between the fission modes. 
In purely adiabatic approaches, the TKE is usually estimated from the scission configuration which is identified on the potential energy surface, based on some criteria \cite{bon06,dub08,sch14a}.
The advantage of using the TDHF+BCS approach is not only to include non-adiabatic effects in the formation of the fragments, but also to provide a well defined value of the TKE. 
Here, the TKE is computed from Coulomb and kinetic energies in post scission configurations following the method described in Ref. \cite{sim14}.
The TKE are found to be 238 MeV, 185 MeV and 163 MeV for the scf, aef, and sef modes, respectively. 
We checked that these results do not depend on the initial configuration in a given valley as long as the initial configuration is not too close to the scission point. 
These results are compared with experimental data \cite{hul89} in Fig. \ref{fig:exp_TKE}. 
The compact symmetric mode is located near the main peak of the TKE distribution.
This observation is in agreement with the experimental data which have attributed this high TKE peak to symmetric fission \cite{hul89}. 
The lower TKE tail is mostly attributed to the asymmetric mode. 
Our calculations also predict that the sef mode leads to a low TKE where only few events have been observed experimentally. 

A particularly interesting feature of quantum microscopic approaches is the possibility to investigate the role of shell effects in the dynamics.
For instance,  TDHF calculations have recently shown the importance of shell effects in the formation of fragments in heavy-ion collisions \cite{wak14,obe14}. 
Here, shell effects  in the tin region (due to the proton magic number $Z=50$) are expected to be present in the symmetric  fission mode of $^{258}$Fm \cite{hul86,hul89}. 
This is compatible with the spherical shape of the fragments in the scf mode (see Fig.~\ref{fig:film}) as well as with the high TKE associated to this mode. 
Indeed, magic fragments are difficult to excite and deform and, thus, fission occurs faster as less dissipation is involved, leading to a larger TKE. 
This is also compatible with the short time associated with the non-adiabatic descent of the potential to fission for the scf mode (see Fig.~\ref{fig:film}). 

\begin{figure}[!ht]   
	\centering\includegraphics[width=\linewidth]{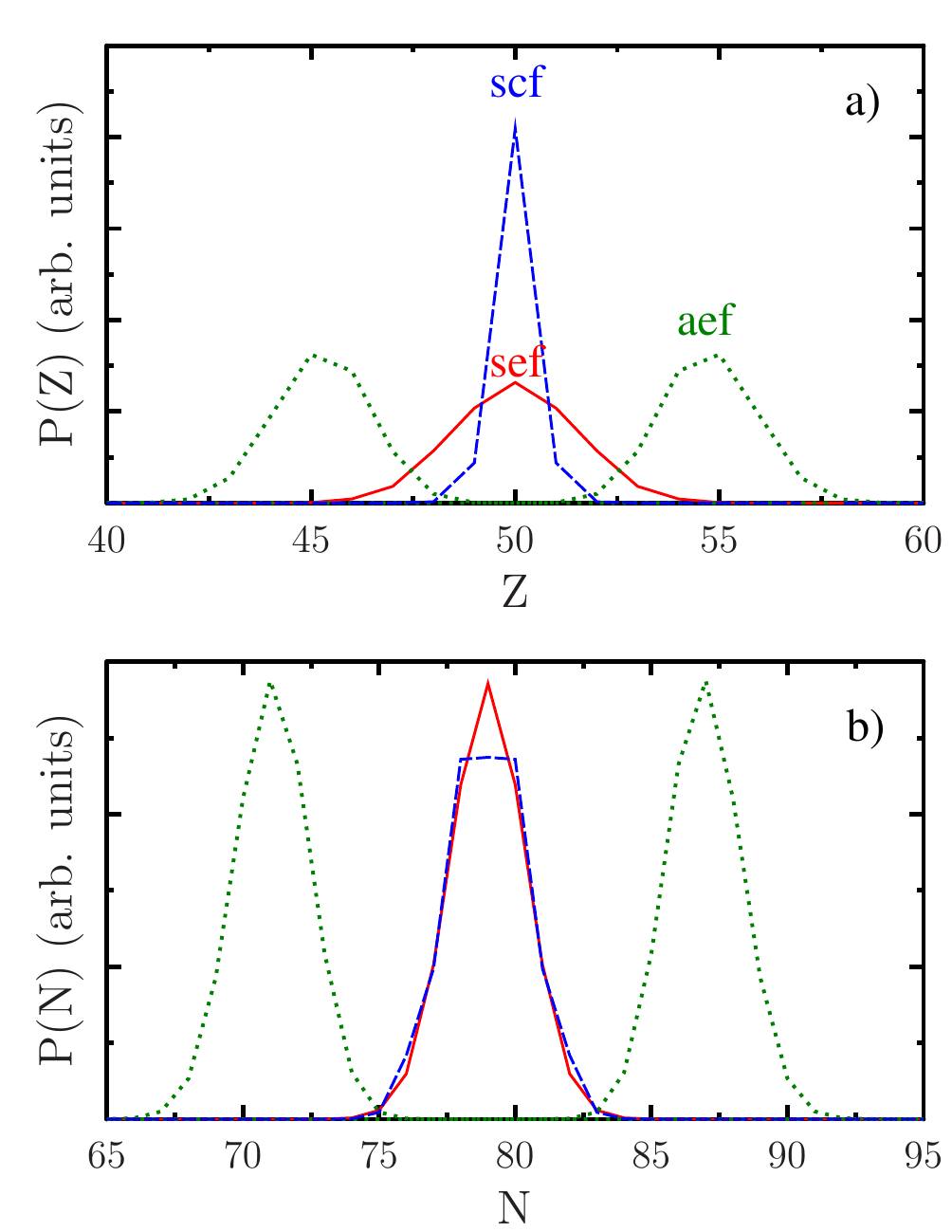}  
	\caption{Proton (a) and neutron (b) number distributions in the fragments for the scf mode (blue dashed line), the aef mode (green doted line) and the sef mode (solid red line).  } 
	\label{fig:prob_3_mode} 
\end{figure}

Another possible signature of shell effects in  $^{258}$Fm fission is the narrow peak in the fragment mass distribution at symmetry (see inset in Fig.~\ref{fig:exp_TKE}) \cite{ich09}. 
In order to see the influence of shell effects on the distributions, we have computed the proton number $Z$ and neutron number $N$ distributions in the fragments at the end of the TDHF+BCS calculations using particle number projection techniques \cite{sim10b,sca13} with the pfaffian calculated using optimized algorithm \cite{Wim12}.
 The resulting distributions are shown in Fig. \ref{fig:prob_3_mode}.
We clearly see that the $Z$ distribution is much sharper for the scf peak than for the other peaks, in good agreement with the expectation that this mode is dominated by spherical shell effects at $Z=50$. 
The peaks in $N$ distributions are all of similar widths, indicating that no shell effect are contributing for neutrons. 
Interestingly enough, we observe a strong odd-even effect in the $N$ distribution of the scf mode due to neutron pairing correlations which is lower in the other modes.

It should be noted that the $Z$ and $N$ distributions shown in Fig.~\ref{fig:prob_3_mode} are not expected to reproduce the widths of the experimental distribution for two reasons. 
The first reason is that these calculations account only for fluctuations acquired during the non-adiabatic phase. 
Fluctuations in the adiabatic phase are indeed expected to be important as shown, e.g., in calculations based on the time-dependent generator coordinate method \cite{gou05} or using the Langevin equation \cite{ran11}. 
The second reason is that the TDHF+BCS approach is expected to underestimate fluctuations of one-body observables \cite{das79,bal81}.
Yet, it would be interesting to have a better estimate of the fluctuations acquired in the non-adiabatic phase.
This can be achieved using beyond mean-field theories such as stochastic approaches \cite{lac14} or the time-dependent random-phase approximation (TDRPA) \cite{bal84}.
Realistic applications of the latter to  nuclear dynamics  have been recently achieved without pairing interaction \cite{bro08,sim11,sim12b}. 
Solving numerically the TDRPA equations with pairing correlations is beyond the scope of this work. 
Nevertheless, we have estimated the enhancement of fluctuations due to beyond-mean-field effects in the non-adiabatic phase of fission for the $^{264}$Fm nucleus fissioning into two $^{132}$Sn doubly magic nuclei in which pairing correlations can be neglected. 
Numerical details for solving the TDRPA equations can be found in \cite{bon85,bro08,sim11,sim12b}. 
As a result, the standard deviation for the distribution of the  total number of nucleons $A$ in the fragment is $\sigma_{TDRPA}=2.35$ for the scf mode, to be compared with the TDHF result $\sigma_{TDHF}=1.35$.
This clearly indicates that beyond mean-field fluctuations also play  an important role in the non-adiabatic phase of fission.

A fully microscopic approach to the fission process has been presented. 
The path to fission is divided into a slow, adiabatic evolution across the fission barrier, followed by a faster, non-adiabatic descent of the potential down to scission described with a time-dependent mean-field approach. 
The method includes pairing correlations and can then be applied to superfluid systems across the nuclear chart. 
Application to the fission of $^{258}$Fm shows a good agreement with experimental data. 
In particular, this approach can be used to determine the total kinetic energy of the fragments without making any assumption on the scission configuration. 
Quantum shell effects are shown to play an important role in the dynamics, in particular in the formation of the fragments in the non-adiabatic phase. 
The method could be generalized to systems with finite temperature in order to study the disappearance of shell effects \cite{mcd14,sch14b}. 
Beyond mean-field fluctuations are shown to be important.
Thus, a quantitative description of fragment distributions requires further developments of beyond mean-field approaches.

We are grateful to M. Bender for his help in performing the calculations in the adiabatic phase. 
D. J. Hinde is thanked for useful comments and discussions. This work has been supported by the
Australian Research Council Grants No. FT120100760.
Part of the calculations have been performed on the NCI National Facility in Canberra, Australia, which is supported
by the Australian Commonwealth Government.
G.S  thanks the JSPS for his JSPS postdoctoral grant for foreign researchers.

\bibliography{biblio}

\begin{thebibliography}{49}
\expandafter\ifx\csname natexlab\endcsname\relax\def\natexlab#1{#1}\fi
\expandafter\ifx\csname bibnamefont\endcsname\relax
  \def\bibnamefont#1{#1}\fi
\expandafter\ifx\csname bibfnamefont\endcsname\relax
  \def\bibfnamefont#1{#1}\fi
\expandafter\ifx\csname citenamefont\endcsname\relax
  \def\citenamefont#1{#1}\fi
\expandafter\ifx\csname url\endcsname\relax
  \def\url#1{\texttt{#1}}\fi
\expandafter\ifx\csname urlprefix\endcsname\relax\def\urlprefix{URL }\fi
\providecommand{\bibinfo}[2]{#2}
\providecommand{\eprint}[2][]{\url{#2}}

\bibitem[{\citenamefont{Hahn and Strassmann}(1939)}]{hah39}
\bibinfo{author}{\bibfnamefont{O.}~\bibnamefont{Hahn}} \bibnamefont{and}
  \bibinfo{author}{\bibfnamefont{F.}~\bibnamefont{Strassmann}},
  \bibinfo{journal}{Naturwissenschaften} \textbf{\bibinfo{volume}{27}},
  \bibinfo{pages}{11} (\bibinfo{year}{1939}), ISSN \bibinfo{issn}{0028-1042},
  \urlprefix\url{http://dx.doi.org/10.1007/BF01488241}.

\bibitem[{\citenamefont{Meitner and Frisch}(1939)}]{mei39}
\bibinfo{author}{\bibfnamefont{L.}~\bibnamefont{Meitner}} \bibnamefont{and}
  \bibinfo{author}{\bibfnamefont{O.~R.} \bibnamefont{Frisch}},
  \bibinfo{journal}{Nature (London)} \textbf{\bibinfo{volume}{143}},
  \bibinfo{pages}{239} (\bibinfo{year}{1939}).

\bibitem[{\citenamefont{Goriely et~al.}(2013)\citenamefont{Goriely, Sida,
  Lema\^{i}tre, Panebianco, Dubray, Hilaire, Bauswein, and Janka}}]{gor13}
\bibinfo{author}{\bibfnamefont{S.}~\bibnamefont{Goriely}},
  \bibinfo{author}{\bibfnamefont{J.-L.} \bibnamefont{Sida}},
  \bibinfo{author}{\bibfnamefont{J.-F.} \bibnamefont{Lema\^{i}tre}},
  \bibinfo{author}{\bibfnamefont{S.}~\bibnamefont{Panebianco}},
  \bibinfo{author}{\bibfnamefont{N.}~\bibnamefont{Dubray}},
  \bibinfo{author}{\bibfnamefont{S.}~\bibnamefont{Hilaire}},
  \bibinfo{author}{\bibfnamefont{A.}~\bibnamefont{Bauswein}}, \bibnamefont{and}
  \bibinfo{author}{\bibfnamefont{H.-T.} \bibnamefont{Janka}},
  \bibinfo{journal}{Phys. Rev. Lett.} \textbf{\bibinfo{volume}{111}},
  \bibinfo{pages}{242502} (\bibinfo{year}{2013}),
  \urlprefix\url{http://link.aps.org/doi/10.1103/PhysRevLett.111.242502}.

\bibitem[{\citenamefont{Moreau and Heyde}(1991)}]{mor91}
\bibinfo{author}{\bibfnamefont{J.}~\bibnamefont{Moreau}} \bibnamefont{and}
  \bibinfo{author}{\bibfnamefont{K.}~\bibnamefont{Heyde}}, in
  \emph{\bibinfo{booktitle}{The Nuclear Fission Process}}, edited by
  \bibinfo{editor}{\bibfnamefont{C.}~\bibnamefont{Wagemans}}
  (\bibinfo{publisher}{CRC Press, Boca Raton, FL}, \bibinfo{year}{1991}).

\bibitem[{\citenamefont{Rizea and Carjan}(2013)}]{riz13}
\bibinfo{author}{\bibfnamefont{M.}~\bibnamefont{Rizea}} \bibnamefont{and}
  \bibinfo{author}{\bibfnamefont{N.}~\bibnamefont{Carjan}},
  \bibinfo{journal}{Nucl. Phys. A} \textbf{\bibinfo{volume}{909}},
  \bibinfo{pages}{50 } (\bibinfo{year}{2013}), ISSN \bibinfo{issn}{0375-9474},
  \urlprefix\url{http://www.sciencedirect.com/science/article/pii/S0375947413005083}.

\bibitem[{\citenamefont{Dubray et~al.}(2008)\citenamefont{Dubray, Goutte, and
  Delaroche}}]{dub08}
\bibinfo{author}{\bibfnamefont{N.}~\bibnamefont{Dubray}},
  \bibinfo{author}{\bibfnamefont{H.}~\bibnamefont{Goutte}}, \bibnamefont{and}
  \bibinfo{author}{\bibfnamefont{J.-P.} \bibnamefont{Delaroche}},
  \bibinfo{journal}{Phys. Rev. C} \textbf{\bibinfo{volume}{77}},
  \bibinfo{pages}{014310} (\bibinfo{year}{2008}),
  \urlprefix\url{http://link.aps.org/doi/10.1103/PhysRevC.77.014310}.

\bibitem[{\citenamefont{Goutte et~al.}(2005)\citenamefont{Goutte, Berger,
  Casoli, and Gogny}}]{gou05}
\bibinfo{author}{\bibfnamefont{H.}~\bibnamefont{Goutte}},
  \bibinfo{author}{\bibfnamefont{J.~F.} \bibnamefont{Berger}},
  \bibinfo{author}{\bibfnamefont{P.}~\bibnamefont{Casoli}}, \bibnamefont{and}
  \bibinfo{author}{\bibfnamefont{D.}~\bibnamefont{Gogny}},
  \bibinfo{journal}{Phys. Rev. C} \textbf{\bibinfo{volume}{71}},
  \bibinfo{pages}{024316} (\bibinfo{year}{2005}),
  \urlprefix\url{http://link.aps.org/doi/10.1103/PhysRevC.71.024316}.

\bibitem[{\citenamefont{Bonneau}(2006)}]{bon06}
\bibinfo{author}{\bibfnamefont{L.}~\bibnamefont{Bonneau}},
  \bibinfo{journal}{Phys. Rev. C} \textbf{\bibinfo{volume}{74}},
  \bibinfo{pages}{014301} (\bibinfo{year}{2006}),
  \urlprefix\url{http://link.aps.org/doi/10.1103/PhysRevC.74.014301}.

\bibitem[{\citenamefont{Staszczak et~al.}(2009)\citenamefont{Staszczak, Baran,
  Dobaczewski, and Nazarewicz}}]{sta09}
\bibinfo{author}{\bibfnamefont{A.}~\bibnamefont{Staszczak}},
  \bibinfo{author}{\bibfnamefont{A.}~\bibnamefont{Baran}},
  \bibinfo{author}{\bibfnamefont{J.}~\bibnamefont{Dobaczewski}},
  \bibnamefont{and}
  \bibinfo{author}{\bibfnamefont{W.}~\bibnamefont{Nazarewicz}},
  \bibinfo{journal}{Phys. Rev. C} \textbf{\bibinfo{volume}{80}},
  \bibinfo{pages}{014309} (\bibinfo{year}{2009}),
  \urlprefix\url{http://link.aps.org/doi/10.1103/PhysRevC.80.014309}.

\bibitem[{\citenamefont{Pei et~al.}(2009)\citenamefont{Pei, Nazarewicz, Sheikh,
  and Kerman}}]{pei09}
\bibinfo{author}{\bibfnamefont{J.~C.} \bibnamefont{Pei}},
  \bibinfo{author}{\bibfnamefont{W.}~\bibnamefont{Nazarewicz}},
  \bibinfo{author}{\bibfnamefont{J.~A.} \bibnamefont{Sheikh}},
  \bibnamefont{and} \bibinfo{author}{\bibfnamefont{A.~K.}
  \bibnamefont{Kerman}}, \bibinfo{journal}{Phys. Rev. Lett.}
  \textbf{\bibinfo{volume}{102}}, \bibinfo{pages}{192501}
  (\bibinfo{year}{2009}),
  \urlprefix\url{http://link.aps.org/doi/10.1103/PhysRevLett.102.192501}.

\bibitem[{\citenamefont{Younes and Gogny}(2011)}]{you11}
\bibinfo{author}{\bibfnamefont{W.}~\bibnamefont{Younes}} \bibnamefont{and}
  \bibinfo{author}{\bibfnamefont{D.}~\bibnamefont{Gogny}},
  \bibinfo{journal}{Phys. Rev. Lett.} \textbf{\bibinfo{volume}{107}},
  \bibinfo{pages}{132501} (\bibinfo{year}{2011}),
  \urlprefix\url{http://link.aps.org/doi/10.1103/PhysRevLett.107.132501}.

\bibitem[{\citenamefont{Warda and Egido}(2012)}]{war12}
\bibinfo{author}{\bibfnamefont{M.}~\bibnamefont{Warda}} \bibnamefont{and}
  \bibinfo{author}{\bibfnamefont{J.~L.} \bibnamefont{Egido}},
  \bibinfo{journal}{Phys. Rev. C} \textbf{\bibinfo{volume}{86}},
  \bibinfo{pages}{014322} (\bibinfo{year}{2012}),
  \urlprefix\url{http://link.aps.org/doi/10.1103/PhysRevC.86.014322}.

\bibitem[{\citenamefont{Abusara et~al.}(2012)\citenamefont{Abusara, Afanasjev,
  and Ring}}]{abu12}
\bibinfo{author}{\bibfnamefont{H.}~\bibnamefont{Abusara}},
  \bibinfo{author}{\bibfnamefont{A.~V.} \bibnamefont{Afanasjev}},
  \bibnamefont{and} \bibinfo{author}{\bibfnamefont{P.}~\bibnamefont{Ring}},
  \bibinfo{journal}{Phys. Rev. C} \textbf{\bibinfo{volume}{85}},
  \bibinfo{pages}{024314} (\bibinfo{year}{2012}),
  \urlprefix\url{http://link.aps.org/doi/10.1103/PhysRevC.85.024314}.

\bibitem[{\citenamefont{Mirea}(2012)}]{mir12}
\bibinfo{author}{\bibfnamefont{M.}~\bibnamefont{Mirea}},
  \bibinfo{journal}{Phys. Lett. B} \textbf{\bibinfo{volume}{717}},
  \bibinfo{pages}{252 } (\bibinfo{year}{2012}), ISSN \bibinfo{issn}{0370-2693},
  \urlprefix\url{http://www.sciencedirect.com/science/article/pii/S0370269312009707}.

\bibitem[{\citenamefont{Staszczak et~al.}(2013)\citenamefont{Staszczak, Baran,
  and Nazarewicz}}]{sta13}
\bibinfo{author}{\bibfnamefont{A.}~\bibnamefont{Staszczak}},
  \bibinfo{author}{\bibfnamefont{A.}~\bibnamefont{Baran}}, \bibnamefont{and}
  \bibinfo{author}{\bibfnamefont{W.}~\bibnamefont{Nazarewicz}},
  \bibinfo{journal}{Phys. Rev. C} \textbf{\bibinfo{volume}{87}},
  \bibinfo{pages}{024320} (\bibinfo{year}{2013}),
  \urlprefix\url{http://link.aps.org/doi/10.1103/PhysRevC.87.024320}.

\bibitem[{\citenamefont{McDonnell et~al.}(2013)\citenamefont{McDonnell,
  Nazarewicz, and Sheikh}}]{mcd13}
\bibinfo{author}{\bibfnamefont{J.~D.} \bibnamefont{McDonnell}},
  \bibinfo{author}{\bibfnamefont{W.}~\bibnamefont{Nazarewicz}},
  \bibnamefont{and} \bibinfo{author}{\bibfnamefont{J.~A.}
  \bibnamefont{Sheikh}}, \bibinfo{journal}{Phys. Rev. C}
  \textbf{\bibinfo{volume}{87}}, \bibinfo{pages}{054327}
  (\bibinfo{year}{2013}),
  \urlprefix\url{http://link.aps.org/doi/10.1103/PhysRevC.87.054327}.

\bibitem[{\citenamefont{Sadhukhan et~al.}(2013)\citenamefont{Sadhukhan,
  Mazurek, Baran, Dobaczewski, Nazarewicz, and Sheikh}}]{sad13}
\bibinfo{author}{\bibfnamefont{J.}~\bibnamefont{Sadhukhan}},
  \bibinfo{author}{\bibfnamefont{K.}~\bibnamefont{Mazurek}},
  \bibinfo{author}{\bibfnamefont{A.}~\bibnamefont{Baran}},
  \bibinfo{author}{\bibfnamefont{J.}~\bibnamefont{Dobaczewski}},
  \bibinfo{author}{\bibfnamefont{W.}~\bibnamefont{Nazarewicz}},
  \bibnamefont{and} \bibinfo{author}{\bibfnamefont{J.~A.}
  \bibnamefont{Sheikh}}, \bibinfo{journal}{Phys. Rev. C}
  \textbf{\bibinfo{volume}{88}}, \bibinfo{pages}{064314}
  (\bibinfo{year}{2013}),
  \urlprefix\url{http://link.aps.org/doi/10.1103/PhysRevC.88.064314}.

\bibitem[{\citenamefont{Simenel and Umar}(2014)}]{sim14}
\bibinfo{author}{\bibfnamefont{C.}~\bibnamefont{Simenel}} \bibnamefont{and}
  \bibinfo{author}{\bibfnamefont{A.~S.} \bibnamefont{Umar}},
  \bibinfo{journal}{Phys. Rev. C} \textbf{\bibinfo{volume}{89}},
  \bibinfo{pages}{031601} (\bibinfo{year}{2014}),
  \urlprefix\url{http://link.aps.org/doi/10.1103/PhysRevC.89.031601}.

\bibitem[{\citenamefont{Schunck et~al.}(2014)\citenamefont{Schunck, Duke, Carr,
  and Knoll}}]{sch14a}
\bibinfo{author}{\bibfnamefont{N.}~\bibnamefont{Schunck}},
  \bibinfo{author}{\bibfnamefont{D.}~\bibnamefont{Duke}},
  \bibinfo{author}{\bibfnamefont{H.}~\bibnamefont{Carr}}, \bibnamefont{and}
  \bibinfo{author}{\bibfnamefont{A.}~\bibnamefont{Knoll}},
  \bibinfo{journal}{Phys. Rev. C} \textbf{\bibinfo{volume}{90}},
  \bibinfo{pages}{054305} (\bibinfo{year}{2014}),
  \urlprefix\url{http://link.aps.org/doi/10.1103/PhysRevC.90.054305}.

\bibitem[{\citenamefont{Negele et~al.}(1978)\citenamefont{Negele, Koonin,
  M\"oller, Nix, and Sierk}}]{neg78}
\bibinfo{author}{\bibfnamefont{J.~W.} \bibnamefont{Negele}},
  \bibinfo{author}{\bibfnamefont{S.~E.} \bibnamefont{Koonin}},
  \bibinfo{author}{\bibfnamefont{P.}~\bibnamefont{M\"oller}},
  \bibinfo{author}{\bibfnamefont{J.~R.} \bibnamefont{Nix}}, \bibnamefont{and}
  \bibinfo{author}{\bibfnamefont{A.~J.} \bibnamefont{Sierk}},
  \bibinfo{journal}{Phys. Rev. C} \textbf{\bibinfo{volume}{17}},
  \bibinfo{pages}{1098} (\bibinfo{year}{1978}),
  \urlprefix\url{http://link.aps.org/doi/10.1103/PhysRevC.17.1098}.

\bibitem[{\citenamefont{Avez et~al.}(2008)\citenamefont{Avez, Simenel, and
  Chomaz}}]{ave08}
\bibinfo{author}{\bibfnamefont{B.}~\bibnamefont{Avez}},
  \bibinfo{author}{\bibfnamefont{C.}~\bibnamefont{Simenel}}, \bibnamefont{and}
  \bibinfo{author}{\bibfnamefont{P.}~\bibnamefont{Chomaz}},
  \bibinfo{journal}{Phys. Rev. C} \textbf{\bibinfo{volume}{78}},
  \bibinfo{eid}{044318} (\bibinfo{year}{2008}).

\bibitem[{\citenamefont{Ebata et~al.}(2010)\citenamefont{Ebata, Nakatsukasa,
  Inakura, Yoshida, Hashimoto, and Yabana}}]{eba10}
\bibinfo{author}{\bibfnamefont{S.}~\bibnamefont{Ebata}},
  \bibinfo{author}{\bibfnamefont{T.}~\bibnamefont{Nakatsukasa}},
  \bibinfo{author}{\bibfnamefont{T.}~\bibnamefont{Inakura}},
  \bibinfo{author}{\bibfnamefont{K.}~\bibnamefont{Yoshida}},
  \bibinfo{author}{\bibfnamefont{Y.}~\bibnamefont{Hashimoto}},
  \bibnamefont{and} \bibinfo{author}{\bibfnamefont{K.}~\bibnamefont{Yabana}},
  \bibinfo{journal}{Phys. Rev. C} \textbf{\bibinfo{volume}{82}},
  \bibinfo{pages}{034306} (\bibinfo{year}{2010}),
  \urlprefix\url{http://link.aps.org/doi/10.1103/PhysRevC.82.034306}.

\bibitem[{\citenamefont{Stetcu et~al.}(2011)\citenamefont{Stetcu, Bulgac,
  Magierski, and Roche}}]{ste11}
\bibinfo{author}{\bibfnamefont{I.}~\bibnamefont{Stetcu}},
  \bibinfo{author}{\bibfnamefont{A.}~\bibnamefont{Bulgac}},
  \bibinfo{author}{\bibfnamefont{P.}~\bibnamefont{Magierski}},
  \bibnamefont{and} \bibinfo{author}{\bibfnamefont{K.~J.} \bibnamefont{Roche}},
  \bibinfo{journal}{Phys. Rev. C} \textbf{\bibinfo{volume}{84}},
  \bibinfo{pages}{051309} (\bibinfo{year}{2011}),
  \urlprefix\url{http://link.aps.org/doi/10.1103/PhysRevC.84.051309}.

\bibitem[{\citenamefont{Hashimoto}(2012)}]{has12}
\bibinfo{author}{\bibfnamefont{Y.}~\bibnamefont{Hashimoto}},
  \bibinfo{journal}{Eur. Phys. J. A} \textbf{\bibinfo{volume}{48}},
  \bibinfo{pages}{55} (\bibinfo{year}{2012}), ISSN \bibinfo{issn}{1434-6001}.

\bibitem[{\citenamefont{Scamps and Lacroix}(2013)}]{sca13}
\bibinfo{author}{\bibfnamefont{G.}~\bibnamefont{Scamps}} \bibnamefont{and}
  \bibinfo{author}{\bibfnamefont{D.}~\bibnamefont{Lacroix}},
  \bibinfo{journal}{Phys. Rev. C} \textbf{\bibinfo{volume}{87}},
  \bibinfo{pages}{014605} (\bibinfo{year}{2013}),
  \urlprefix\url{http://link.aps.org/doi/10.1103/PhysRevC.87.014605}.

\bibitem[{\citenamefont{Hulet et~al.}(1989)\citenamefont{Hulet, Wild, Dougan,
  Lougheed, Landrum, Dougan, Baisden, Henderson, Dupzyk, Hahn et~al.}}]{hul89}
\bibinfo{author}{\bibfnamefont{E.~K.} \bibnamefont{Hulet}},
  \bibinfo{author}{\bibfnamefont{J.~F.} \bibnamefont{Wild}},
  \bibinfo{author}{\bibfnamefont{R.~J.} \bibnamefont{Dougan}},
  \bibinfo{author}{\bibfnamefont{R.~W.} \bibnamefont{Lougheed}},
  \bibinfo{author}{\bibfnamefont{J.~H.} \bibnamefont{Landrum}},
  \bibinfo{author}{\bibfnamefont{A.~D.} \bibnamefont{Dougan}},
  \bibinfo{author}{\bibfnamefont{P.~A.} \bibnamefont{Baisden}},
  \bibinfo{author}{\bibfnamefont{C.~M.} \bibnamefont{Henderson}},
  \bibinfo{author}{\bibfnamefont{R.~J.} \bibnamefont{Dupzyk}},
  \bibinfo{author}{\bibfnamefont{R.~L.} \bibnamefont{Hahn}},
  \bibnamefont{et~al.}, \bibinfo{journal}{Phys. Rev. C}
  \textbf{\bibinfo{volume}{40}}, \bibinfo{pages}{770} (\bibinfo{year}{1989}),
  \urlprefix\url{http://link.aps.org/doi/10.1103/PhysRevC.40.770}.

\bibitem[{\citenamefont{M\"oller and Iwamoto}(2000)}]{mol00}
\bibinfo{author}{\bibfnamefont{P.}~\bibnamefont{M\"oller}} \bibnamefont{and}
  \bibinfo{author}{\bibfnamefont{A.}~\bibnamefont{Iwamoto}},
  \bibinfo{journal}{Phys. Rev. C} \textbf{\bibinfo{volume}{61}},
  \bibinfo{pages}{047602} (\bibinfo{year}{2000}),
  \urlprefix\url{http://link.aps.org/doi/10.1103/PhysRevC.61.047602}.

\bibitem[{\citenamefont{Ichikawa et~al.}(2009)\citenamefont{Ichikawa, Iwamoto,
  and M\"oller}}]{ich09}
\bibinfo{author}{\bibfnamefont{T.}~\bibnamefont{Ichikawa}},
  \bibinfo{author}{\bibfnamefont{A.}~\bibnamefont{Iwamoto}}, \bibnamefont{and}
  \bibinfo{author}{\bibfnamefont{P.}~\bibnamefont{M\"oller}},
  \bibinfo{journal}{Phys. Rev. C} \textbf{\bibinfo{volume}{79}},
  \bibinfo{pages}{014305} (\bibinfo{year}{2009}),
  \urlprefix\url{http://link.aps.org/doi/10.1103/PhysRevC.79.014305}.

\bibitem[{\citenamefont{Kim et~al.}(1997)\citenamefont{Kim, Otsuka, and
  Bonche}}]{kim97}
\bibinfo{author}{\bibfnamefont{K.-H.} \bibnamefont{Kim}},
  \bibinfo{author}{\bibfnamefont{T.}~\bibnamefont{Otsuka}}, \bibnamefont{and}
  \bibinfo{author}{\bibfnamefont{P.}~\bibnamefont{Bonche}},
  \bibinfo{journal}{J. Phys. G} \textbf{\bibinfo{volume}{23}},
  \bibinfo{pages}{1267} (\bibinfo{year}{1997}).

\bibitem[{\citenamefont{Bonche et~al.}(2005)\citenamefont{Bonche, Flocard, and
  Heenen}}]{bon05}
\bibinfo{author}{\bibfnamefont{P.}~\bibnamefont{Bonche}},
  \bibinfo{author}{\bibfnamefont{H.}~\bibnamefont{Flocard}}, \bibnamefont{and}
  \bibinfo{author}{\bibfnamefont{P.~H.} \bibnamefont{Heenen}},
  \bibinfo{journal}{Comp. Phys. Com.} \textbf{\bibinfo{volume}{171}},
  \bibinfo{pages}{49} (\bibinfo{year}{2005}), ISSN \bibinfo{issn}{0010-4655},
  \urlprefix\url{http://www.sciencedirect.com/science/article/pii/S0010465505002821}.

\bibitem[{\citenamefont{B{\l}ocki and Flocard}(1976)}]{blo76}
\bibinfo{author}{\bibfnamefont{J.}~\bibnamefont{B{\l}ocki}} \bibnamefont{and}
  \bibinfo{author}{\bibfnamefont{H.}~\bibnamefont{Flocard}},
  \bibinfo{journal}{Nucl. Phys. A} \textbf{\bibinfo{volume}{273}},
  \bibinfo{pages}{45 } (\bibinfo{year}{1976}), ISSN \bibinfo{issn}{0375-9474},
  \urlprefix\url{http://www.sciencedirect.com/science/article/pii/0375947476902992}.

\bibitem[{\citenamefont{Maruhn et~al.}(2005)\citenamefont{Maruhn, Reinhard,
  Stevenson, Stone, and Strayer}}]{Mar05}
\bibinfo{author}{\bibfnamefont{J.~A.} \bibnamefont{Maruhn}},
  \bibinfo{author}{\bibfnamefont{P.~G.} \bibnamefont{Reinhard}},
  \bibinfo{author}{\bibfnamefont{P.~D.} \bibnamefont{Stevenson}},
  \bibinfo{author}{\bibfnamefont{J.~R.} \bibnamefont{Stone}}, \bibnamefont{and}
  \bibinfo{author}{\bibfnamefont{M.~R.} \bibnamefont{Strayer}},
  \bibinfo{journal}{Phys. Rev. C} \textbf{\bibinfo{volume}{71}},
  \bibinfo{eid}{064328} (\bibinfo{year}{2005}).

\bibitem[{\citenamefont{Goddard}(2014)}]{god14}
\bibinfo{author}{\bibfnamefont{P.~M.} \bibnamefont{Goddard}}, Ph.D. thesis,
  \bibinfo{school}{University of Surrey} (\bibinfo{year}{2014}),
  \urlprefix\url{http://epubs.surrey.ac.uk/id/eprint/806610}.

\bibitem[{\citenamefont{Wakhle et~al.}(2014)\citenamefont{Wakhle, Simenel,
  Hinde, Dasgupta, Evers, Luong, du~Rietz, and Williams}}]{wak14}
\bibinfo{author}{\bibfnamefont{A.}~\bibnamefont{Wakhle}},
  \bibinfo{author}{\bibfnamefont{C.}~\bibnamefont{Simenel}},
  \bibinfo{author}{\bibfnamefont{D.~J.} \bibnamefont{Hinde}},
  \bibinfo{author}{\bibfnamefont{M.}~\bibnamefont{Dasgupta}},
  \bibinfo{author}{\bibfnamefont{M.}~\bibnamefont{Evers}},
  \bibinfo{author}{\bibfnamefont{D.~H.} \bibnamefont{Luong}},
  \bibinfo{author}{\bibfnamefont{R.}~\bibnamefont{du~Rietz}}, \bibnamefont{and}
  \bibinfo{author}{\bibfnamefont{E.}~\bibnamefont{Williams}},
  \bibinfo{journal}{Phys. Rev. Lett.} \textbf{\bibinfo{volume}{113}},
  \bibinfo{pages}{182502} (\bibinfo{year}{2014}),
  \urlprefix\url{http://link.aps.org/doi/10.1103/PhysRevLett.113.182502}.

\bibitem[{\citenamefont{Oberacker et~al.}(2014)\citenamefont{Oberacker, Umar,
  and Simenel}}]{obe14}
\bibinfo{author}{\bibfnamefont{V.~E.} \bibnamefont{Oberacker}},
  \bibinfo{author}{\bibfnamefont{A.~S.} \bibnamefont{Umar}}, \bibnamefont{and}
  \bibinfo{author}{\bibfnamefont{C.}~\bibnamefont{Simenel}},
  \bibinfo{journal}{Phys. Rev. C} \textbf{\bibinfo{volume}{90}},
  \bibinfo{pages}{054605} (\bibinfo{year}{2014}),
  \urlprefix\url{http://link.aps.org/doi/10.1103/PhysRevC.90.054605}.

\bibitem[{\citenamefont{Hulet et~al.}(1986)\citenamefont{Hulet, Wild, Dougan,
  Lougheed, Landrum, Dougan, Schadel, Hahn, Baisden, Henderson et~al.}}]{hul86}
\bibinfo{author}{\bibfnamefont{E.~K.} \bibnamefont{Hulet}},
  \bibinfo{author}{\bibfnamefont{J.~F.} \bibnamefont{Wild}},
  \bibinfo{author}{\bibfnamefont{R.~J.} \bibnamefont{Dougan}},
  \bibinfo{author}{\bibfnamefont{R.~W.} \bibnamefont{Lougheed}},
  \bibinfo{author}{\bibfnamefont{J.~H.} \bibnamefont{Landrum}},
  \bibinfo{author}{\bibfnamefont{A.~D.} \bibnamefont{Dougan}},
  \bibinfo{author}{\bibfnamefont{M.}~\bibnamefont{Schadel}},
  \bibinfo{author}{\bibfnamefont{R.~L.} \bibnamefont{Hahn}},
  \bibinfo{author}{\bibfnamefont{P.~A.} \bibnamefont{Baisden}},
  \bibinfo{author}{\bibfnamefont{C.~M.} \bibnamefont{Henderson}},
  \bibnamefont{et~al.}, \bibinfo{journal}{Phys. Rev. Lett.}
  \textbf{\bibinfo{volume}{56}}, \bibinfo{pages}{313} (\bibinfo{year}{1986}),
  \urlprefix\url{http://link.aps.org/doi/10.1103/PhysRevLett.56.313}.

\bibitem[{\citenamefont{Simenel}(2010)}]{sim10b}
\bibinfo{author}{\bibfnamefont{C.}~\bibnamefont{Simenel}},
  \bibinfo{journal}{Phys. Rev. Lett.} \textbf{\bibinfo{volume}{105}},
  \bibinfo{pages}{192701} (\bibinfo{year}{2010}),
  \urlprefix\url{http://link.aps.org/doi/10.1103/PhysRevLett.105.192701}.

\bibitem[{\citenamefont{Wimmer}(2012)}]{Wim12}
\bibinfo{author}{\bibfnamefont{M.}~\bibnamefont{Wimmer}}, \bibinfo{journal}{ACM
  Trans. Math. Softw.} \textbf{\bibinfo{volume}{38}}, \bibinfo{pages}{30:1}
  (\bibinfo{year}{2012}), ISSN \bibinfo{issn}{0098-3500},
  \urlprefix\url{http://doi.acm.org/10.1145/2331130.2331138}.

\bibitem[{\citenamefont{Randrup and M\"oller}(2011)}]{ran11}
\bibinfo{author}{\bibfnamefont{J.}~\bibnamefont{Randrup}} \bibnamefont{and}
  \bibinfo{author}{\bibfnamefont{P.}~\bibnamefont{M\"oller}},
  \bibinfo{journal}{Phys. Rev. Lett.} \textbf{\bibinfo{volume}{106}},
  \bibinfo{pages}{132503} (\bibinfo{year}{2011}),
  \urlprefix\url{http://link.aps.org/doi/10.1103/PhysRevLett.106.132503}.

\bibitem[{\citenamefont{Dasso et~al.}(1979)\citenamefont{Dasso, D{\o}ssing, and
  Pauli}}]{das79}
\bibinfo{author}{\bibfnamefont{C.~H.} \bibnamefont{Dasso}},
  \bibinfo{author}{\bibfnamefont{T.}~\bibnamefont{D{\o}ssing}},
  \bibnamefont{and} \bibinfo{author}{\bibfnamefont{H.~C.} \bibnamefont{Pauli}},
  \bibinfo{journal}{Z. Phys. A} \textbf{\bibinfo{volume}{289}},
  \bibinfo{pages}{395} (\bibinfo{year}{1979}).

\bibitem[{\citenamefont{Balian and V\'en\'eroni}(1981)}]{bal81}
\bibinfo{author}{\bibfnamefont{R.}~\bibnamefont{Balian}} \bibnamefont{and}
  \bibinfo{author}{\bibfnamefont{M.}~\bibnamefont{V\'en\'eroni}},
  \bibinfo{journal}{Phys. Rev. Lett.} \textbf{\bibinfo{volume}{47}},
  \bibinfo{pages}{1353} (\bibinfo{year}{1981}),
  \urlprefix\url{http://link.aps.org/doi/10.1103/PhysRevLett.47.1353}.

\bibitem[{\citenamefont{Lacroix and Ayik}(2014)}]{lac14}
\bibinfo{author}{\bibfnamefont{D.}~\bibnamefont{Lacroix}} \bibnamefont{and}
  \bibinfo{author}{\bibfnamefont{S.}~\bibnamefont{Ayik}},
  \bibinfo{journal}{Eur. Phys. J. A} \textbf{\bibinfo{volume}{50}},
  \bibinfo{eid}{95} (\bibinfo{year}{2014}), ISSN \bibinfo{issn}{1434-6001},
  \urlprefix\url{http://dx.doi.org/10.1140/epja/i2014-14095-8}.

\bibitem[{\citenamefont{Balian and V\'en\'eroni}(1984)}]{bal84}
\bibinfo{author}{\bibfnamefont{R.}~\bibnamefont{Balian}} \bibnamefont{and}
  \bibinfo{author}{\bibfnamefont{M.}~\bibnamefont{V\'en\'eroni}},
  \bibinfo{journal}{Phys. Lett. B} \textbf{\bibinfo{volume}{136}},
  \bibinfo{pages}{301} (\bibinfo{year}{1984}), ISSN \bibinfo{issn}{0370-2693},
  \urlprefix\url{http://www.sciencedirect.com/science/article/pii/0370269384920082}.

\bibitem[{\citenamefont{Broomfield and Stevenson}(2008)}]{bro08}
\bibinfo{author}{\bibfnamefont{J.~M.~A.} \bibnamefont{Broomfield}}
  \bibnamefont{and} \bibinfo{author}{\bibfnamefont{P.~D.}
  \bibnamefont{Stevenson}}, \bibinfo{journal}{J. Phys. G}
  \textbf{\bibinfo{volume}{35}}, \bibinfo{pages}{095102}
  (\bibinfo{year}{2008}).

\bibitem[{\citenamefont{Simenel}(2011)}]{sim11}
\bibinfo{author}{\bibfnamefont{C.}~\bibnamefont{Simenel}},
  \bibinfo{journal}{Phys. Rev. Lett.} \textbf{\bibinfo{volume}{106}},
  \bibinfo{pages}{112502} (\bibinfo{year}{2011}),
  \urlprefix\url{http://link.aps.org/doi/10.1103/PhysRevLett.106.112502}.

\bibitem[{\citenamefont{Simenel}(2012)}]{sim12b}
\bibinfo{author}{\bibfnamefont{C.}~\bibnamefont{Simenel}},
  \bibinfo{journal}{Eur. Phys. J. A} \textbf{\bibinfo{volume}{48}},
  \bibinfo{pages}{152} (\bibinfo{year}{2012}), ISSN \bibinfo{issn}{1434-6001},
  \urlprefix\url{http://dx.doi.org/10.1140/epja/i2012-12152-0}.

\bibitem[{\citenamefont{Bonche and Flocard}(1985)}]{bon85}
\bibinfo{author}{\bibfnamefont{P.}~\bibnamefont{Bonche}} \bibnamefont{and}
  \bibinfo{author}{\bibfnamefont{H.}~\bibnamefont{Flocard}},
  \bibinfo{journal}{Nucl. Phys. A} \textbf{\bibinfo{volume}{437}},
  \bibinfo{pages}{189} (\bibinfo{year}{1985}), ISSN \bibinfo{issn}{0375-9474},
  \urlprefix\url{http://www.sciencedirect.com/science/article/pii/0375947485902325}.

\bibitem[{\citenamefont{McDonnell et~al.}(2014)\citenamefont{McDonnell,
  Nazarewicz, Sheikh, Staszczak, and Warda}}]{mcd14}
\bibinfo{author}{\bibfnamefont{J.~D.} \bibnamefont{McDonnell}},
  \bibinfo{author}{\bibfnamefont{W.}~\bibnamefont{Nazarewicz}},
  \bibinfo{author}{\bibfnamefont{J.~A.} \bibnamefont{Sheikh}},
  \bibinfo{author}{\bibfnamefont{A.}~\bibnamefont{Staszczak}},
  \bibnamefont{and} \bibinfo{author}{\bibfnamefont{M.}~\bibnamefont{Warda}},
  \bibinfo{journal}{Phys. Rev. C} \textbf{\bibinfo{volume}{90}},
  \bibinfo{pages}{021302} (\bibinfo{year}{2014}),
  \urlprefix\url{http://link.aps.org/doi/10.1103/PhysRevC.90.021302}.

\bibitem[{\citenamefont{Schunck et~al.}()\citenamefont{Schunck, Duke, and
  Carr}}]{sch14b}
\bibinfo{author}{\bibfnamefont{N.}~\bibnamefont{Schunck}},
  \bibinfo{author}{\bibfnamefont{D.}~\bibnamefont{Duke}}, \bibnamefont{and}
  \bibinfo{author}{\bibfnamefont{H.}~\bibnamefont{Carr}},
  \bibinfo{note}{arXiv:1311.2620}.

\end{thebibliography}

\end{document}